\documentclass[11pt]{article}
\usepackage{moriond,epsfig}
\usepackage{xspace}

\def\be{\begin{equation}}
\def\ee{\end{equation}}
\def\bea{\begin{eqnarray}}
\def\eea{\end{eqnarray}}

\newcommand{\cdf}{CDF\xspace}
\newcommand{\dzero}{D0\xspace}
\newcommand{\ttbar}{\ensuremath{t\bar{t}}\xspace}
\newcommand{\ttbarX}[1]{\ensuremath{t\bar{t}#1}\xspace}

\newcommand{\invfb}{\ensuremath{\mbox{fb}^{-1}}\xspace}
\newcommand{\tZq}{\ensuremath{t\rightarrow Zq}\xspace}
\newcommand{\chisq}{\ensuremath{\chi^{2}}\xspace}
\newcommand{\Ht}{\ensuremath{H_{T}}\xspace}

\newcommand{\unit}[1]{\ensuremath{\mathrm{\:#1}}}

\begin{document}
\vspace*{4cm}
\title{SEARCHES FOR NEW PHYSICS IN TOP EVENTS AT THE TEVATRON}

\author{R. EUSEBI}

\address{Fermi National Accelerator Laboratory,\\
Batavia, Illinois 60510, USA}

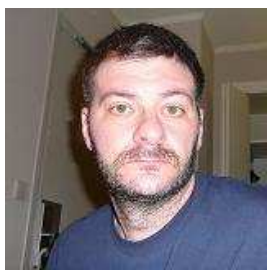
\maketitle
\abstracts{
During the past years the \cdf and \dzero detectors have collected large amounts of data obtaining a relatively pure sample of pair-produced top quarks and a well understood sample containing singly-produced top quarks. These samples have been used for the precise measurement of the top quark properties, and have set stringent limits on new physics in the top sample.  This reports presents the latest results from the \cdf and \dzero collaborations on the search for new physics within the top sample using an integrated data sample of up to 3.6\unit{\invfb}.}

\section{Introduction}
The standard model of particles of fields (SM) predicted the existence of the top quark, a weak isospin partner of the bottom quark.  In 1995 the top quark was discovered by the \cdf and \dzero collaborations\cite{topCDF,topDzero} with a mass of about 175\unit{GeV}. Its large mass suggest it is strongly associated with the mechanism of electro-weak symmetry breaking, and makes it the fermion with the largest coupling to the SM-expected, but not yet found, Higgs boson. These reasons make the top quark potentially sensitive to new physics, which can be revealed through both, precision measurements of its production and decay properties and by dedicated searches for new physics in the top quark sample.
This letter reports the results of measurements of top quark properties with up to 3.6\unit{\invfb} of data. In general most of the analyses in this report were performed by both the \cdf and \dzero collaborations, however, for each search a single analysis of either collaboration is presented here.

\section{Searches for New Physics in the Top sample}
\label{newphysics}

\subsection{Top Anomalous Decays}
\label{topano}
In the SM the top is expected to decay to a $W$ boson and a $b$ quark with a branching ratio larger than 99\%. The presence of physics beyond the SM, like flavor changing neutral currents (FCNC), can alter this branching ratio. In the SM flavor changing neutral currents (FCNC) are allowed at orders higher than tree level. The decay \tZq for example is very rare with a branching ratio of about B($\tZq$)$\approx10^{-14}$ in the SM, but with the potential to reach values as high as $10^{-2}$ in exotic scenarios involving new physics\cite{XM_TZQ}.

Based on the number of expected event in the $\ttbar \rightarrow lepton+jets$ channel and using 1.9\unit{\invfb} of data the \cdf collaboration has set limits on the top decays to either $Zc$, $\gamma c$, $gc$ or to an invisible final state. This analysis requires missing transverse energy greater than 30 GeV, at least three jets with energy greater than 30 GeV, and at least two loose $b$-tags, see~\cite{invi} for details.
By comparing the observed event yields with the predicted ones this analysis set upper limits at the 95\% confidence level (C.L.) of $BR(t\rightarrow Z c)<0.15$, $BR(t\rightarrow g c)<0.14$, $BR(t\rightarrow \gamma c)<0.12$, and $BR(t\rightarrow invisible)<0.10$.

\subsection{Stringent Limits on Flavor Changing Neutral Currents}
Unlike the general search shown above the \cdf collaboration has performed a targeted search for the flavor changing neutral current decay of the top quark \tZq using a data sample corresponding to an integrated luminosity of  1.9\unit{\invfb}. Candidate events are selected by requiring two opposite sign leptons ($e$'s or $\mu$'s), 4 or more jets and a series of optimized cuts.  Events in this signal region are further classified according to whether or not they have a secondary vertex, or $b$-tag. A third sample is used as control and made from rejected events that failed to pass at least one of the optimized requirements.

 The signal is discriminated from the background by exploring kinematic constraints present in FCNC  events. A mass \chisq variable quantifies the consistency of each event with originating from a top quark FCNC decay.  Templates of this variable are generated for the main backgrounds, and the FCNC signal. Shape systematic uncertainties are included in the templates.
The \chisq template fit is implemented as a simultaneous fit to two signal regions and the control region. Assuming  a top quark mass of $175$\unit{GeV} the expect sensitivity of the measurement is to set an upper limit on $B(\tZq)$ of $5.0\%$. The results of the fit are consistent with the \chisq distribution of the background. An upper limit of $B(\tZq)< 3.7\%$ at 95\% C.L. is obtained using the Feldman-Cousins prescription, which is much more stringent that the one obtained by comparing yields as shown in the previous section.

\subsection{Top Couplings Form Factors}
The most general $tbW$ coupling including operators up to dimension five can be written as:
\begin{equation}
\mathcal{L} = -\frac{g}{\sqrt{2}}\bar{b}\gamma^{\mu}V_{tb}(f^L_1 P_L + f^R_1 P_R)tW^-_\mu \\
-\frac{g}{\sqrt{2}}\bar{b} \frac{i\sigma^{\mu\nu}q_\nu}{M_W} (f_2^L P_L + f_2^R P_R)W^-_\mu + h.c.
\end{equation}
where $M_W$ is the mass of the W boson, $q_\nu$ is its four-momentum, $V^{tb}$ is the Cabibbo-Kobayashi-Maskawa matrix element, $P_L = (1-\gamma^5)/2$ $(P_R = (1+\gamma^5)/2)$ is the left-handed (right-handed) projection operator and $f_1^L$, $f_1^R$, $f_2^L$ and $f_2^R$ are the form factors couplings, which in the SM take the values $f^L_1=1, f^L_2 = f^R_1 = f^R_2 = 0$. The lower indices 1 and 2 of the couplings refer to the vector and tensor characters of the coupling respectively.

While measurements of the  BR($b \rightarrow s\gamma)$ have been used to constrain the right-handed vector and tensor couplings~\cite{bsgamma} those measurements rely on assumptions of the absence of other 15 non-SM contributions to the $b$ quark decay. The direct measurement presented here avoids such assumptions.
The \dzero collaboration has set limits on all this factors using up to 2.7\unit{\invfb} of data. 

This analysis investigate one pair of coupling form factors at a time out of the full set of left/right (upper indices L or R) and vector/tensor (lower indices 1 or 2) form factors. For each pair of couplings under investigation we assume that the other two couplings take their SM values.

Using a measurement of the helicity of the W bosons in the decay of $\ttbar \rightarrow lepton+jets$ events with 2.7\unit{\invfb} of data~\cite{WHEL} a likelihood is obtained in the ($f_1^L$,$f_1^R$), ($f_1^L$,$f_2^L$) and ($f_1^L$,$f_2^R$) planes. This likelihood is combined with the result of a search for anomalous couplings in the single top quark final state using 1.0\unit{\invfb} of data. The result of the combination is a two-dimensional posterior probability density as a function of both form factors in each of the three two-dimensional planes.

By projecting the two-dimensional posteriors in the corresponding form factor axis we set 95\% C.L limits of $|f_1^L|^2$ $|f_1^R|^2<0.72$, $|f_2^L|^2<0.3$, and  $|f_2^R|^2<0.19$. The most stringent limit of  $|f_1^L|^{2}<1.16^{+0.51}_{-0.44}$ is obtained from the posterior ($f_1^L$,$f_2^R$). See full details at~\cite{anomCoup}.

\subsection{Pair production of Stop Quarks}

The theory of super-symmetry proposes that each particle in the SM has a corresponding super-partner. In this theory the super-symmetric partner of the top quark is a scalar particle called the stop quark.
The \cdf collaboration has searched for pair-produced stop quarks in which each stop decays via $\tilde{t}_1 \rightarrow b \tilde{\chi}_1^{\pm} \rightarrow b\tilde{\chi}^0_1 l \nu$, providing a signature similar to that of the SM $\ttbar\rightarrow dilepton$ process. The analysis examines 2.7\unit{\invfb} of data selecting events with two leptons, two jets, and missing transverse energy to account for the undetected particles in the stop decay chains. A special cut is imposed to reduce background from Drell-Yan processes. This analysis is separated into two channels; events that contain a $b$-tagged jet, and events that do not.

The events are kinematically reconstructed under the stop decay hypothesis. The reconstructed stop mass is used as a discriminating kinematic variable in a fit to data. The reconstructed stop mass distribution depends on the mass of the stop ($M(\tilde{t})$), the mass of the neutralino ($M(\chi^0)$) and on the branching ratio of the chargino to a lepton and neutralino (BR($\bar{\chi}^{\pm}_1 \rightarrow \tilde{\chi}^0 \nu l$)). This analysis set limits at 95\% C.L. on the ($M(\tilde{t}),M(\chi^0)$) plane for different values of BR($\bar{\chi}^{\pm}_1 \rightarrow \tilde{\chi}^0 \nu l$). See~\cite{stops} to see the two dimensional exclusion counturs and for details on the analysis.

\subsection{Search for 4th Generation Top}
Fourth generation $t'$'s are predicted in some SUSY models~\cite{SMFOUR}. The \cdf collaboration has searched for a heavy top ($t'$) quark pair production decaying to $Wq$ final states in 2.8\unit{\invfb} in the lepton plus jets data sample without b-tagging requirements.  The $t'$ is assumed to be produced in pairs via the strong interaction, to have mass greater than the top quark, and to decays promptly and only to $Wq$ final states.

Two variables are directly related to the mass of $t'$; the total transverse energy of the event (\Ht), and the reconstructed mass of the $t'$ ($M_{reco}$) as obtained from a kinematic fitter. To discriminate the new physics signal from standard model backgrounds a set of 2D-templates of the main backgrounds, as well as different mass $t'$'s, are constructed in the (\Ht,$M_{reco}$) plane. For a given $t'$ mass, the observed data is fitted to the background 2D-template and to the 2D-template of the given $t'$ mass, to set limits on the $t'$ production. Using a specific $t'$ model~\cite{XM_TPRIME2}, this analysis exclude $t'$ with masses below 311\unit{GeV/C^2} at 95\% C.L.

\subsection{Higgs Production in Association with \ttbar}
The \dzero collaboration has searched for Higgs production in association with \ttbar production using 2.1\unit{\invfb} of data. Data is selected with a single lepton and four or more jets, with one to three $b$-tagged jets.  For each event $H_T$, the scalar sum of the transverse energies of the lepton and all jets, is computed and its distribution in all jet-multiplicity bins is used to set limits on the $\ttbarX{H}\rightarrow \ttbarX{b\bar{b}}$ production cross section.

Kinematical differences between \ttbar and \ttbarX{H} events are exploited and limits are set on the $\ttbarX{H}\rightarrow \ttbarX{b\bar{b}}$ production cross section for different Higgs masses. This analysis finds limits at 95\% C.L. that range from 48 times the SM-expected cross section for a Higgs mass of 105\unit{GeV}, to about 835 times the SM-expected cross section for a Higgs mass of 155\unit{GeV} . See all the details of this analysis at~\cite{ttbarH}.

These limits are also interpreted as limits on the production of $\bar{t} t'$ where the $t'$ is a fourth generation top that decays via $t' \rightarrow t H$. This kind of production is detailed at~\cite{ttbarH_interpret}. The results of this interpretation are expressed as exclusion limits in the two dimensional plane of ($M_H,M_{t'}$). See~\cite{ttbarH} for details on this analysis.

\subsection{Top Quark Pair Resonances}
The \dzero collaboration has performed a search of a resonance ($X$) decaying to a pair of top quarks. This analysis requires events with a single lepton, large missing transverse energy and at least three jets selecting  3.6\unit{\invfb} of data.

The distribution of the events total invariant mass observed in data is compared to templates for the expected SM backgrounds and narrow resonance signals of various masses.  Since no excess is seen above backgrounds limits are obtained on $\sigma_{X}\times$ BR($X\rightarrow \ttbar$), where $\sigma_{X}$ is the production cross section of the resonance and BR($X\rightarrow \ttbar$) represents the branching ratio of the decay of the resonance to \ttbar.  Within a top-color-assisted Technicolor model, the existence of a leptophobic $Z'$ boson with $M_{Z'}<$820\unit{GeV} and width $\Gamma_{Z'}=0.012 M_{Z'}$ is excluded at 95\% C.L. See~\cite{ttbarreso} for details.

\section*{Acknowledgments}
I would like to thank the \cdf and \dzero collaborations for the excellent quality of their work, in particular  the authors of all the analyses shown here for their critical input, as well as top group conveners of the respective collaborations for their support and advice. I also thank the organizers of the Moriond QCD 2009 conference.

\end{document}